\documentclass[%
 amsmath,amssymb,
prb,twocolumn
]{revtex4-1}

\usepackage[dvipdfmx]{graphicx}
\usepackage{dcolumn}
\usepackage{bm}
\usepackage{color}

\renewcommand{\section}[1]{\textit{#1.---}} 
\renewcommand{\vec}[1]{\boldsymbol{#1}}

\begin{document}

\title{Tight-binding theory of surface spin states on bismuth thin films}
\author{Kazuo Saito, Hirokatsu Sawahata, Takashi Komine and Tomosuke Aono}
\affiliation{ 
Faculty of Engineering, Ibaraki University\\
Hitachi 316-8511, Japan}
\email{aono@mx.ibaraki.ac.jp}
\date{\today}

\begin{abstract}
The surface spin states for bismuth thin films are investigated
using an $sp^3$ tight-binding model.
The model explains most experimental observations using 
angle-resolved photoemission spectroscopy,
including the Fermi surface, the band structure with
Rashba spin splitting, and the quantum confinement in the energy band gap of
the surface states.
A large out-of-plane spin component also appears.
The surface states penetrate inside the film to within approximately a few bilayers near
the Brillouin-zone center, whereas they reach the center of the film near
the Brillouin-zone boundary.
\end{abstract}
\maketitle

\section{Introduction} \label{sec:}
The spin-orbit interaction (SOI) induces spin splitting
in the absence of an external magnetic field
on a two-dimensional (2D) system,
i.e.,
Rashba spin splitting~\cite{Bychkov:1984aa},
which has been an indispensable element of spintronic physics and devices~\cite{Winkler:2003aa}.
The Rashba effect is expected on crystal surfaces
due to their inversion asymmetry.
For example, Rashba spin splitting has been observed on the Au(111) surface~\cite{LaShell:1996aa,Nicolay:2001aa,Henk:2004aa}.
Bismuth (Bi) is a group V semimetal with a large SOI due to the heavy mass of the Bi atom; therefore,
the surface of Bi crystals is an ideal system to
observe a strong Rashba effect~\cite{Hofmann:2006aa}.

Angle-resolved photoemission spectroscopy (APRES) experiments have been reported for the Bi surface accompanied with first-principles band calculations~\cite{ %
Ast:2001aa,
Koroteev:2004aa,
Kim:2005aa,
Hirahara:2006aa,
Hirahara:2007ac,
Hirahara:2007aa,
Hirahara:2008aa,
Kimura:2010aa,
Takayama:2011aa,
Ohtsubo:2012aa,
Takayama:2012aa,
Takayama:2014aa,
Takayama:2015aa}.
The surface states have
a hexagonal electron pocket around the $\bar{\Gamma}$ point
and six-fold hole pockets
~\cite{ %
Ast:2001aa,
Kim:2005aa,
Hirahara:2006aa,
Ohtsubo:2012aa,
Takayama:2015aa}
First-principles band calculations showed that
these two surface states are spin-split bands
~\cite{%
Koroteev:2004aa,
Hirahara:2006aa},
 and this Rashba splitting
has been confirmed experimentally
~\cite { %
Hirahara:2007aa,
Hirahara:2008aa,
Kimura:2010aa}.
In addition,
the surface spin orientation has been elucidated, and in particular, 
a giant out-of-plane spin polarization was reported
~\cite{Takayama:2011aa}.
The band structure is dependent on the film thickness because of
the quantum confinement effect~\cite{Hirahara:2006aa,Hirahara:2007aa,Takayama:2012aa}.

In addition to the ARPES experiments,
many interesting features have been studied in
the electronic transport properties of Bi nanostructures.
The conductivity of Bi films has been measured
~\cite{ %
Hirahara:2007ab,
Jnawali:2012aa,
Aitani:2014aa},
and was determined to be
dependent on both the surface and bulk states,
and the coupling between them has a major influence on the conductivity
Quantum confinement effects can significantly enhance thermoelectric properties
in quantum well and quantum wire structures~\cite{Hicks:1993aa,*Hicks:1993ab}. 
Bismuth is thus a prime candidate to achieve high performance thermoelectric conversion in its nanostructures~\cite{Hicks:1993ac,Murata:2009aa}.
To understand the transport properties in Bi quantum confinement structures,
it is necessary to simultaneously determine the electronic properties of
both the surface and the bulk states.

Although first-principles band calculations
have already revealed the Fermi surface and the energy band structure
~\cite{%
Koroteev:2004aa,
Hirahara:2006aa,
Hirahara:2007aa,
Koroteev:2008aa},
no systematic analysis 
for comparison with the reported ARPES experimental results 
has been conducted to date.
Here, we approach this issue using an $sp^3$ tight-binding model
that reproduces the band structure of bulk Bi proposed by Liu and Allen~\cite{Liu:1995aa}.
This model has been applied to discuss the topological and non-topological phases of the surface states of pure Bi and Sb~\cite{Fukui:2007aa}, and Bi$_{1-x}$Sb$_{x}$~\cite{Teo:2008aa}, 
as well as two-dimensional Bi~\cite{Murakami:2006aa}.
Extra surface hopping terms
~\cite{Petersen:2000aa,Ast:2012ab}
are added that were originally proposed to explain
the Au(111) surface states.
This model will enable confirmation of
whether the ARPES results originate from the surface effect.
In addition, it is straightforward to see the effects of quantum confinement
because the film thickness can be easily changed
and the electronic states both inside the film and at the surface can be analyzed,
which is important to investigate the electronic transport properties.
We can thus give a systematic survey of
the ARPES experimental results by taking advantage of these points.

\section{Model Hamiltonian} \label{sec:}
\begin{figure*}[htb]
\centering
\includegraphics[width=\textwidth]{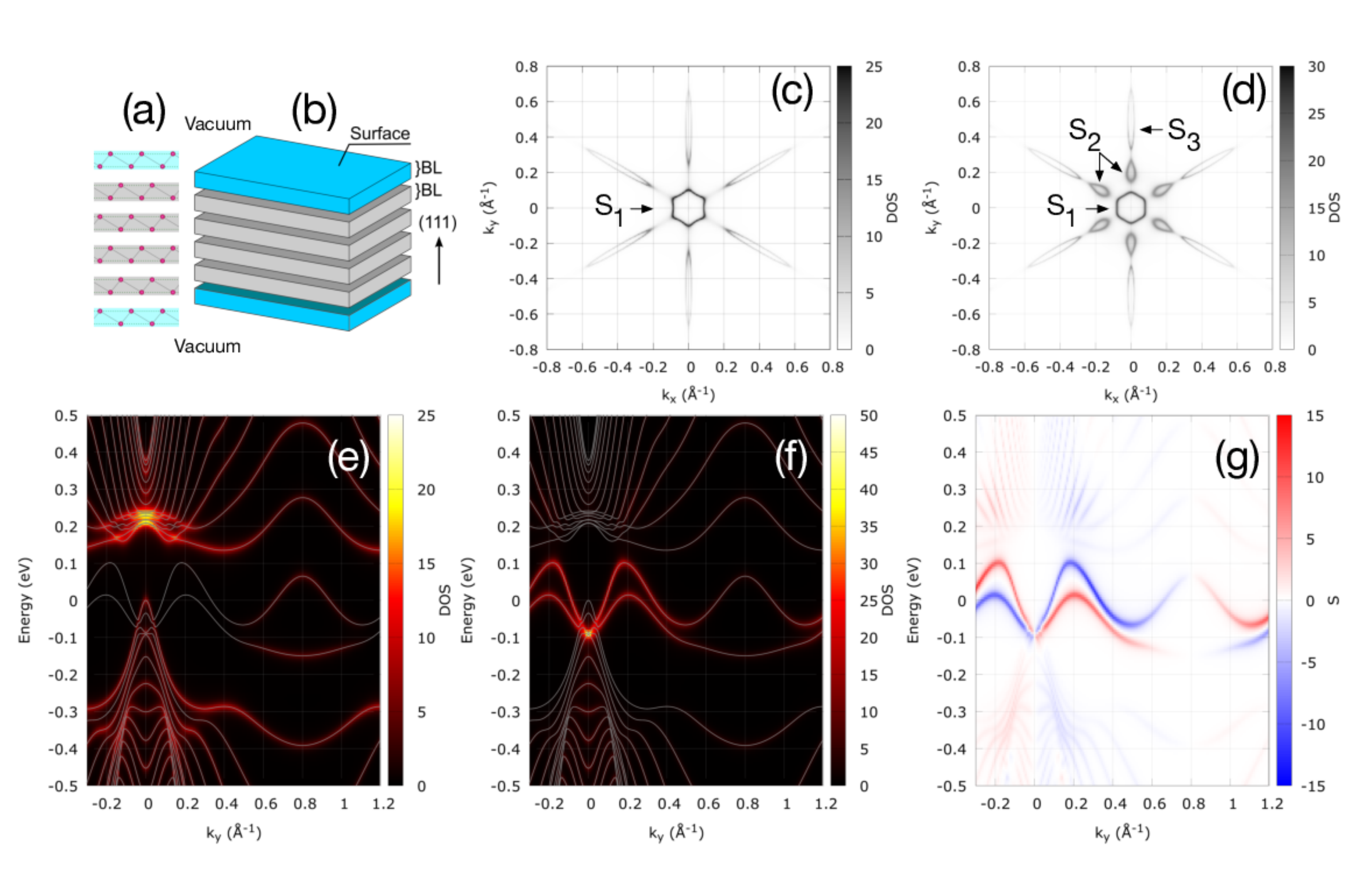}
\caption{%
(Color online)
(a) Bismuth crystal structure.
(b) Schematic view of Bi thin film.
(c, d) Surface BL density of states (DOS) at the Fermi energy $\rho(1,0,k_x,k_y)$ of a 16 BL film
with (c)
$\gamma_{\textrm{pp/sp}}=0$,
and (d)
with $\gamma_{\textrm{sp}}=0.45$ and $\gamma_{\textrm{pp}}=-0.27$.
(e, f) Energy band structure along the $\bar{\Gamma}$-$\bar{\textrm{M}}$ line
($k_x = 0$).
 The DOS $\rho(i,E,0,k_y)$ is plotted
for the (e) middle BL ($i=8$) and (f) the surface BL ($i=1$).
The white lines represent the eigenvalues of $H$.
(g) Spin-resolved band structure for (f),
where $s=s_x(1,E,0,k_y)$. Note that $s_{y/z}(1,E,0,k_y)=0$.
}
\label{fig:surface DOS and Band}
\end{figure*}

Bismuth has a rhombohedral Bravais lattice with two atoms per unit cell, forming a bilayer (BL) structure,
as shown in Fig.~\ref{fig:surface DOS and Band}(a).
A Bi thin film is obtained by stacking the BLs along the (111) direction,
such as the $z$-axis depicted in Fig.~\ref{fig:surface DOS and Band}(b).
The surface is thus parallel to the $xy$ plane.
The uppermost and lowermost BLs
are in contact with a vacuum.

We first construct a model Hamiltonian for the Bi thin film.
For this purpose, the $sp^3$ tight-binding model proposed for the bulk Bi crystal~\cite{Liu:1995aa} is adapted to the Bi thin film.
There are
$s$-, $p_x$-, $p_y$-, and $p_z$-orbitals with spin index $\sigma$ on each atom.
The hopping terms among the atomic orbitals are
decomposed into inter- and intra-BL hopping terms.
The inter-BL hopping term $H_{21\mathchar`-2}$
consists of the nearest-neighbor hopping term
in the bulk Bi Hamiltonian, whereas
the intra-BL hopping term consists of
two parts, $H_{11}$ and $H_{12\mathchar`-1}$,
with the third and second nearest-neighbor hopping terms in the bulk model, respectively.
The Fermi energy is set to zero.

There is a surface potential gradient on the surface BL
along the $z$-axis
between the surface Bi atoms and the vacuum.
The surface Rashba effect is
induced by the contribution of this potential gradient~\cite{Petersen:2000aa,Ast:2012ab}.
In terms of the $sp^3$ tight-binding model,
this is described by
the following spin independent hopping terms
between the nearest-neighbors sites, $\vec{R}_i$ and $\vec{R}_j$
~\cite{Petersen:2000aa,Ast:2012ab}:
\begin{equation} \label{eq:surface-hopping-term}
t_{\alpha\beta}
=
\begin{cases}
\gamma_{\textrm{pp}}  \cos \theta_{ij} &,
 (\alpha,\beta)=(p_x,p_z) \textrm{\;or\;} (p_z,p_x),\\
\gamma_{\textrm{pp}}  \sin \theta_{ij}&,
 (\alpha,\beta)=(p_y,p_z) \textrm{\;or\;} (p_z,p_y), \\
\gamma_{\textrm{sp}}  &,
 (\alpha,\beta)=(s,p_z) \textrm{\;or\;} (p_z,s),
\end{cases}
\end{equation}
where $\theta_{ij}$ is the azimuthal angle between
$\vec{R}_i - \vec{R}_j$ and the $x$-axis,
and
$\gamma_{\textrm{pp}}$ and $\gamma_{\textrm{sp}}$
are the hopping matrix elements of the Hamiltonian.
Note that those hopping terms are zero in the bulk Bi crystal model 
because of the inversion symmetry.
It is assumed that the surface hopping terms (\ref{eq:surface-hopping-term}) appear
only on the uppermost atomic layer, the first atomic layer of the uppermost BL,
and the lowermost atomic layer
with $-\gamma_{\textrm{sp/pp}}$, 
because the surface field points 
in the opposite direction at the lowermost layer.
The values of $\gamma_{\textrm{sp}}$ and $\gamma_{\textrm{pp}}$ remain to be determined at this stage.

The total Hamiltonian of the thin film $H$ is therefore
represented by the following matrix form: 
\begin{equation}\label{eq:hamiltonian-matrix}
H = 
\begin{pmatrix} 
H_{\textrm{s} 11} & H_{12\mathchar`-1} &   &  &  &  &    \cr
H_{21\mathchar`-1}& H_{11} & H_{21\mathchar`-2} &  &  &  &    \cr
&H_{12\mathchar`-2}& H_{11} & H_{12\mathchar`-1} &  &  &     \cr
&  &H_{21\mathchar`-1} & H_{11} &  &  &     \cr
&  &  & \ddots& \ddots & \ddots &     \cr
& & & & \phantom{H_{12\mathchar`-2}} & H_{11} & H_{12\mathchar`-1}   \cr
& & & & & H_{21\mathchar`-1}& H_{\textrm{s}11}'  \cr
\end{pmatrix},
\end{equation}
where $H_{\textrm{s}11}$ is the Hamiltonian for the uppermost atomic layer,
which includes
the surface hopping terms~(\ref{eq:surface-hopping-term})
in addition to $H_{11}$,
while $H_{\textrm{s}11}'$ is the Hamiltonian
that includes the surface hopping terms (\ref{eq:surface-hopping-term}) with 
$-\gamma_{\textrm{pp/sp}}$.
The size of the matrix is 
thus $16n \times 16n$ when
the number of the BLs is $n$.
The Hamiltonian (\ref{eq:hamiltonian-matrix}) is a function of
the wave vectors $k_x$ and $k_y$:
$H = H(k_x,k_y)$.

\section{Calculation of DOS} \label{sec:}
The DOS and the band structure of the thin film
is obtained from the retarded Green's function matrix
$G(E,k_x,k_y)$ with energy $E$ defined by
\begin{equation} \label{eq:}
G(E,k_x,k_y) = \left[ E + i \delta - H(k_x, k_y) \right]^{-1}
\end{equation}
with $\delta = 1.0\times 10^{-2}$ in the numerical calculations.
The DOS in the $i$th BL is defined by 
\begin{equation} \label{eq:surface-DOS-Green}
\rho(i,E,k_x,k_y)  = -\frac{1}{\pi} \mathrm{Tr}\, \mathrm{Im}\,  G(E,k_x,k_y),
\end{equation}
where $\mathrm{Tr}$ represents the trace over
the orbitals and the spin only on the $i$th BL ($i=1$ for the uppermost BL).
In a similar way,
the spin-resolved DOS $ s_\alpha(i,E,k_x,k_y) \; (\alpha=x,y,z)$ is given by
\begin{equation} \label{eq:surface-SDPS-Green}
s_\alpha(i,E,k_x,k_y) =
-\frac{1}{\pi} \mathrm{Tr}\, \mathrm{Im}\, s_\alpha  G(E,k_x,k_y),
\end{equation}
where $s_\alpha$ is the Pauli spin matrix that acts on the four orbital states. 
The eigenvalues of $H$ are also calculated to show
the entire band structure of the film.

\section{Parameter fitting} \label{sec:}
In the following,
$\gamma_{\textrm{sp}}$ and $\gamma_{\textrm{pp}}$ are treated as
fitting parameters.
To fix these values, we use a phenomenological approach:
We first calculate the DOS on the surface BL and
the band structure for various values of $\gamma_{\textrm{sp/pp}}$ 
and then compare them with the ARPES experimental results
~\cite{ %
Ast:2001aa,
Koroteev:2004aa,
Hirahara:2006aa,
Hirahara:2007ac,
Ohtsubo:2012aa,
Takayama:2015aa}
to find the best selection.
This scheme was successful and led to
$\gamma_{\textrm{sp}}=0.45$, and $\gamma_{\textrm{pp}}=-0.27$.
The numerical results appear similar near these values.
Note that these values are the same order of magnitude as the hopping matrix elements between
the second and third nearest neighbors given in Ref.~\cite{Liu:1995aa}.

The presence of the surface terms (\ref{eq:surface-hopping-term})
is essential to explain the observed Fermi surface.
Figures \ref{fig:surface DOS and Band}(c) and (d),
show the DOS on the surface BL at the Fermi energy $\rho(1,0,k_x,k_y)$
for the 16 BL thin film
without and with the surface hopping terms, respectively.
In both cases,
a hexagonal electron pocket appears around 
the $\bar{\Gamma}$ point designated by $S_1$. 
Qualitative differences arise outside of $S_1$;
with the surface hopping term, there are six hole lobes and six extra electron lobes,
designated by $S_2$, and $S_3$, respectively,
while $S_2$ is missing without the surface hopping term.
The ARPES experiments
show the presence of $S_2$,
which confirms that the surface terms (\ref{eq:surface-hopping-term}) play
a central role in the formation of the Fermi surface.

\section{Band structure} \label{sec:}
Next, we discuss the energy band structure
along the $\bar{\Gamma}$-$\bar{\textrm{M}}$ line ($k_x=0$).
Figures~\ref{fig:surface DOS and Band}(e) and (f)
show $\rho(i,E,0,k_y)$
for the middle ($i=8$) and surface ($i=1$) BLs, respectively.
The eigenvalues of $H$
are also shown as white lines for comparison.
On the middle BL,
the plot covers 
most of the eigenvalues,
while on the surface BL,
the plot appears only in a small fraction of the eigenvalue curves and
mostly on two curves near the Fermi energy.
The upper curve forms the $S_1$ and $S_3$ structures,
whereas
the lower curve forms the $S_2$ structure.

The spin-resolved band structure illustrates
the distinctive features of the surface states,
as shown in Fig~\ref{fig:surface DOS and Band}(g).
The spin splitting appears near the $\bar{\Gamma}$ point,
which is similar to Rashba spin splitting,
and it diminishes near $\bar{\textrm{M}}$.
This is consistent with
the experimental results and the first-principles band calculations~\cite{Hirahara:2007aa,Kimura:2010aa}.
Thus, the surface states on the Bi film are 
well described by the phenomenological tight-binding model.

\begin{figure}[htb]
\centering
\includegraphics[width=0.5\textwidth]{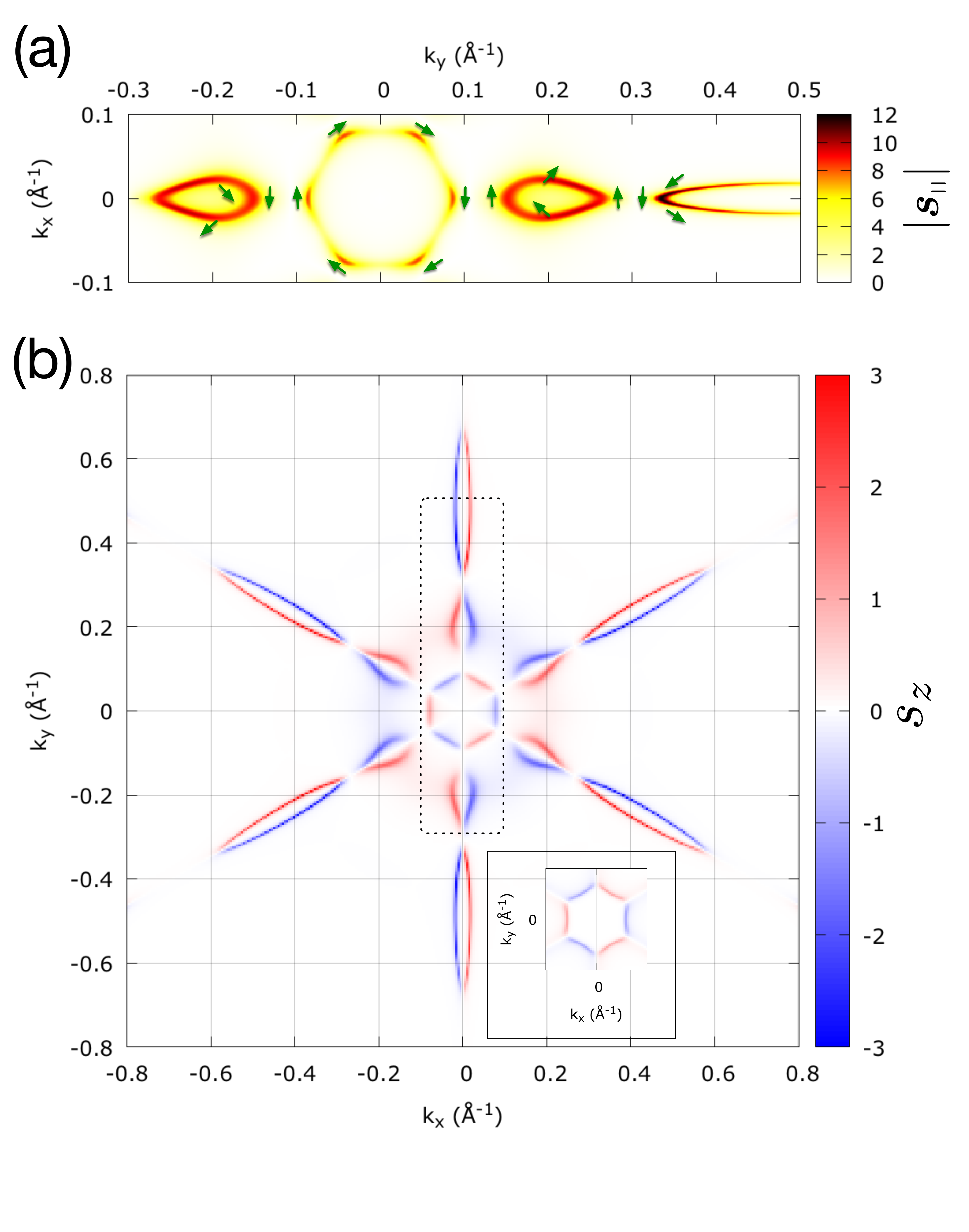}
\caption{%
(Color online)
(a) In-plane surface spin magnitude $|\vec{s_{\shortparallel}}|$. 
The green arrows indicate the direction of $\vec{s_{\shortparallel}}$ for representative points.
(b) Out-of-plane surface spin $s_z$.
The region surrounded by a dotted line corresponds to that shown in (a).
Inset: $s_z$ around the 
$\bar{\Gamma}$ point without the surface hopping terms and with the same scale as the main figure.
 Additional information on the surface spin states is
 given in the Supplemental Material~\cite{supple_bismuth_film}.}
\label{fig:spin-texture}
\end{figure}
\section{Surface spin states}
Next, we discuss the surface spin texture at the Fermi energy;
$s_\alpha \equiv  s_\alpha(1,0,k_x,k_y)$.
Figure~\ref{fig:spin-texture}(a) shows
the in-plane spin $\vec{s_{\shortparallel}}=(s_x,s_y)$ distribution.
On $S_1$,
$\vec{s_{\shortparallel}}$ lies along the pocket structure, 
while on $S_2$,
the direction of $\vec{s}$ is opposite to
that on $S_1$.
The in-plane spin rotations on $S_1$ and $S_2$
are broadly similar to those by the Rashba SOI.
In addition,
$|\vec{s_{\shortparallel}}|$ along the lobe on $S_2$ is almost constant.
These observations are consistent with previous experimental 
results~\cite{Kim:2005aa,Hirahara:2008aa,Takayama:2011aa,Takayama:2014aa}.
However, the asymmetry of the $|\vec{s}_{\shortparallel}|$
along the $k_y$ axis on $S_2$~\cite{Takayama:2011aa} is not observed in the present model.
Instead of this asymmetry,
$|\vec{s_{\shortparallel}}|$ on $S_1$ oscillates every 60$^{\circ}$.
In addition, $\vec{s}$ is not always perpendicular to $\vec{k}$ on $S_2$,
which comes from the non-parabolic band structure.
These indicate that the spin structure is not described by a simple Rashba SOI model.

\begin{figure*}[htb]
\centering
\includegraphics[width=0.7\textwidth]{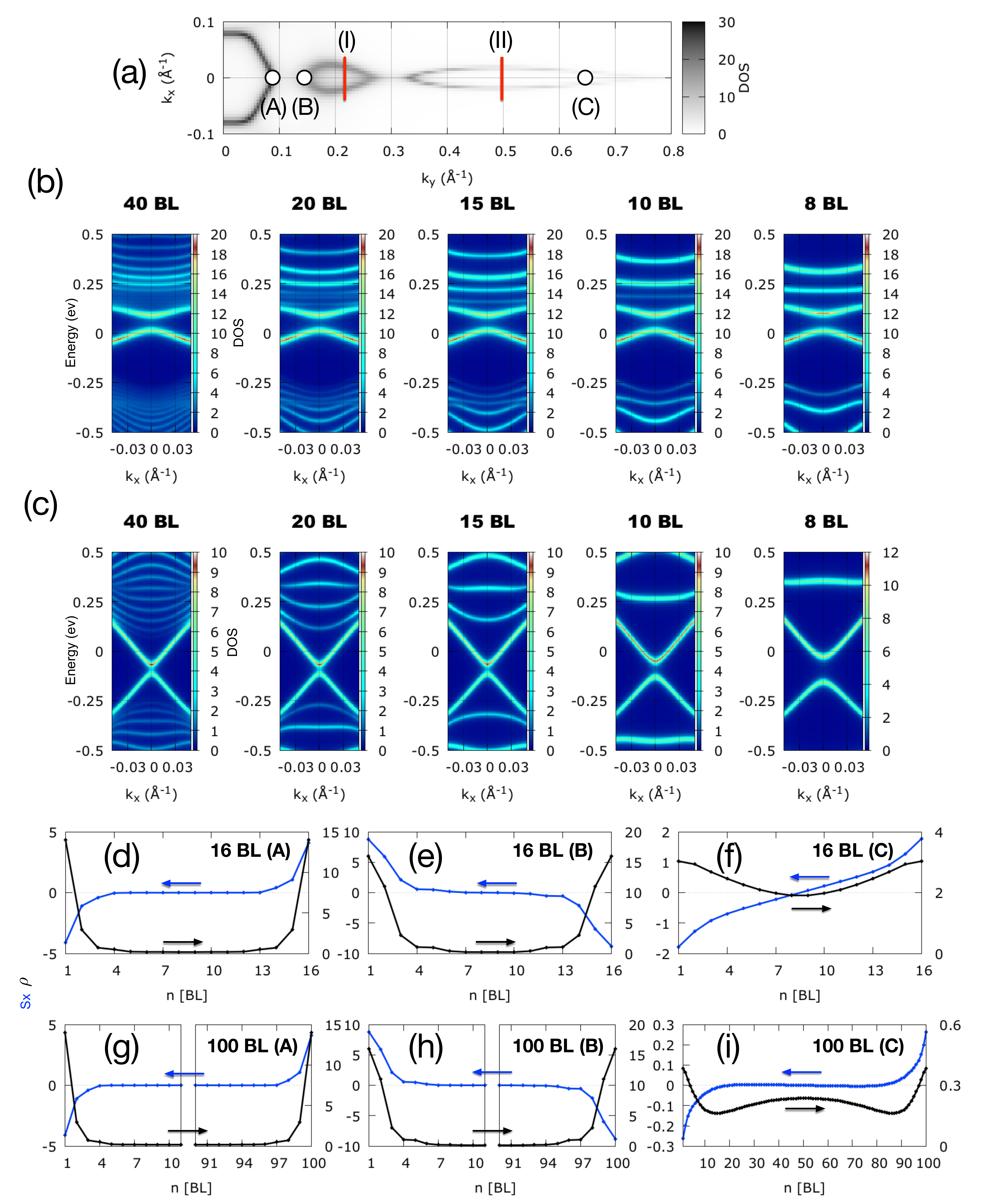}
\caption{%
(Color online)
(a) Surface BL DOS at the Fermi energy $\rho(1,0,k_x,k_y)$ of a 40 BL film
with two lines and three points.
(b, c) Energy band structure for various numbers of BLs.
The surface BL DOS $\rho(1,E,k_x,k_y)$ is plotted
along the lines (I) ($k_y=0.22$) and (II) ($k_y=0.5$), respectively.
(d-i) The layer-resolved DOS at the Fermi energy, $\rho(n)$ and $s_x(n)$,
at $k_y=0.08$, 0.16, and 0.65,
the points (A), (B), and (C) in (a), respectively, 
for (d-f) 16 BL, and (g-i) 100 BL.
In (g) and (h), $\rho(n)$ and $s_x(n)$ are shown only near the surface BLs. }
\label{fig:BL dependence}
\end{figure*}
The deviation from the simple Rashba model is clarified
by the out-of-plane spin $s_z$, 
as shown in Fig.~\ref{fig:spin-texture}(b).
There is a relatively large $s_z$ over the $S_1$--$S_3$ structures, 
where the maximum of $|s_z|$ is approximately 25$\%$ of the maximum of $|\vec{s_{\shortparallel}}|$.
Furthermore,
$s_z$ changes its sign every 60$^{\circ}$.
These results are consistent with recent experimental results~\cite{Takayama:2011aa},
although larger values of $|s_z|$ are observed experimentally.
In addition, the fine structure of $s_z$ is clarified, where 
the sign of $s_z$ also changes from $S_1$ to $S_3$
in the same manner as $\vec{s_{\shortparallel}}$.

We further discuss the presence of the giant $s_z$ component.
The first-principles calculations show
similar results for $s_z$
with the topological phase of the Bi$_{1-x}$Sb$_{x}$ crystal~\cite{Zhang:2009aa},
where $s_z$ is very small and
around 1\% of $|\vec{s_{\shortparallel}}|$,
which indicates 
the spin lies on the two-dimensional surface.
Pure Bi ($x=0$) is in the trivial phase~\cite{Fukui:2007aa,Teo:2008aa,Zhang:2009aa}; therefore,
large values of $s_z$ may be direct evidence for
clarification of the difference between the trivial and topological phases,
besides 
the number of Fermi surface crossings from
the zone center to the boundary.
To support this point within the proposed model,
$s_z$ without the surface hopping term near the $\bar{\Gamma}$ point 
is shown in the inset of Fig.~\ref{fig:spin-texture}(b).
Similar results are obtained both with and without the surface hopping terms. 
This indicates that the origin of $s_z$ is not from
the surface effect, but from
the bulk hopping terms and the atomic SOI of Bi itself,
which determines the bulk band structure.
Thus, the origin of $s_z$ for pure Bi is associated with the bulk band structure,
which leads to a trivial phase.

\section{BL number dependence} \label{sec:}
Figure~\ref{fig:BL dependence}(a) shows $\rho(1,0,k_x,k_y)$ for a 40 BL film.
Compared with that for the 16 BL film,
both $S_1$ and $S_2$ structures are unchanged, while the $S_3$ structure is prolonged
towards the $\bar{M}$ point,
which is consistent with the experimental results~\cite{Hirahara:2006aa,Hirahara:2007ac}.
To examine this difference in detail, we discuss 
the BL number dependence along the two lines and three points shown in Fig.~\ref{fig:BL dependence}(a).

Figure~\ref{fig:BL dependence}(b) shows
$\rho(1,E,k_x,k_y)$ in $S_2$ for various numbers of BLs 
along the line (I) shown in Fig.~\ref{fig:BL dependence}(a).
The two surface states near the Fermi energy
separated by a band gap are not
affected by changing the BL number.
However, the energy levels away from the Fermi level are
under the strong influence of the BL number, which indicates the quantum confinement in the thin film.
Figure~\ref{fig:BL dependence}(c) shows
$\rho(1,E,k_x,k_y)$
in $S_3$ along the line (II) shown in Fig.~\ref{fig:BL dependence}(a).
Although the band structure near the Fermi level shows a linear dispersion
similar to that for line (I),
the band gap clearly decreases as the BL number increases.
A similar observation is obtained experimentally~\cite{Takayama:2012aa}.
Hence, the surface states are under the strong influence of quantum confinement
on $S_3$, while they are not on $S_2$.

Finally
we discuss the surface state penetration inside the thin film.
For this purpose,
the layer-resolved DOS,
$\rho(n) \equiv \rho(n,0,0,k_y)$ and $s_x(n) \equiv s_x(n,0,0,k_y)$, are shown in
Figs.~\ref{fig:BL dependence}(d-f) and (g-i)
for the 16 BL and 100 BL films, respectively,
at the three points indicated in Fig.~\ref{fig:BL dependence}(a).
All the figures show
that 
the spin on the uppermost and lowermost BLs are
in opposite directions, as expected;
The spin changes its sign at the middle of the film.
The surface states on $S_1$ penetrate only a few BLs,
and a similar result is obtained for the surface states on $S_2$
with a slightly longer penetration length.
The penetration length is unchanged by the film thickness,
which confirms they are genuine surface states.
On other hand, at $S_3$,
$\rho(n)$ and $s_x(n)$ decay over 20 BLs,
and $\rho(n)$ is finite even at the middle of the film.
Thus, the states are no longer simple ``surface" states and are under
the influence of the quantum confinement inside the film.

\section{Conclusions} \label{sec:}
We have shown that an $sp^3$ tight-binding model
with surface hopping terms can explain most of the experimental ARPES observations for bismuth thin films,
including the Fermi surface, the spin-resolved band structure with
Rashba spin splitting,
and the quantum confinement effect in the energy band structure.
The model also explains the large out-of-plane spin observed,
which originates from the intrinsic Bi crystal structure rather than
the surface effect.
We have also clearly shown that
the surface states penetrate inside the film to within approximately a few BLs near
the Brillouin-zone center, whereas they reach the center of the film near
the Brillouin-zone boundary.

 
The authors acknowledge A.~Takayama for fruitful discussions and comments
on the ARPES experimental results.
We also thank J.~Ieda for stimulating discussion.
This work is partially supported by a 
Kakenhi Grant-in-Aid from the Ministry of Education, Culture, Sports, Science and Technology (MEXT) of Japan, and the Japan Science and Technology Agency (JST).

%

\pagebreak
\widetext
\begin{center}
\textbf{\large Supplemental Material \\
Tight-binding theory of surface spin states on bismuth thin films}
\end{center}
\setcounter{equation}{0}
\setcounter{figure}{0}
\setcounter{table}{0}
\setcounter{page}{1}
\makeatletter
\renewcommand{\theequation}{S\arabic{equation}}
\renewcommand{\thefigure}{S\arabic{figure}}
\renewcommand{\bibnumfmt}[1]{[S#1]}
\renewcommand{\citenumfont}[1]{S#1}


This supplemental material contains
\begin{itemize}
\item Magnitude and direction of the in-plane spin,
\item Surface spin density of states (DOS) at the Fermi energy.
\end{itemize}

\vspace{1cm}

\section{Magnitude and direction of the in-plane spin} \label{sec:}
\begin{figure}[htb]
\centering
\includegraphics[width=\textwidth]{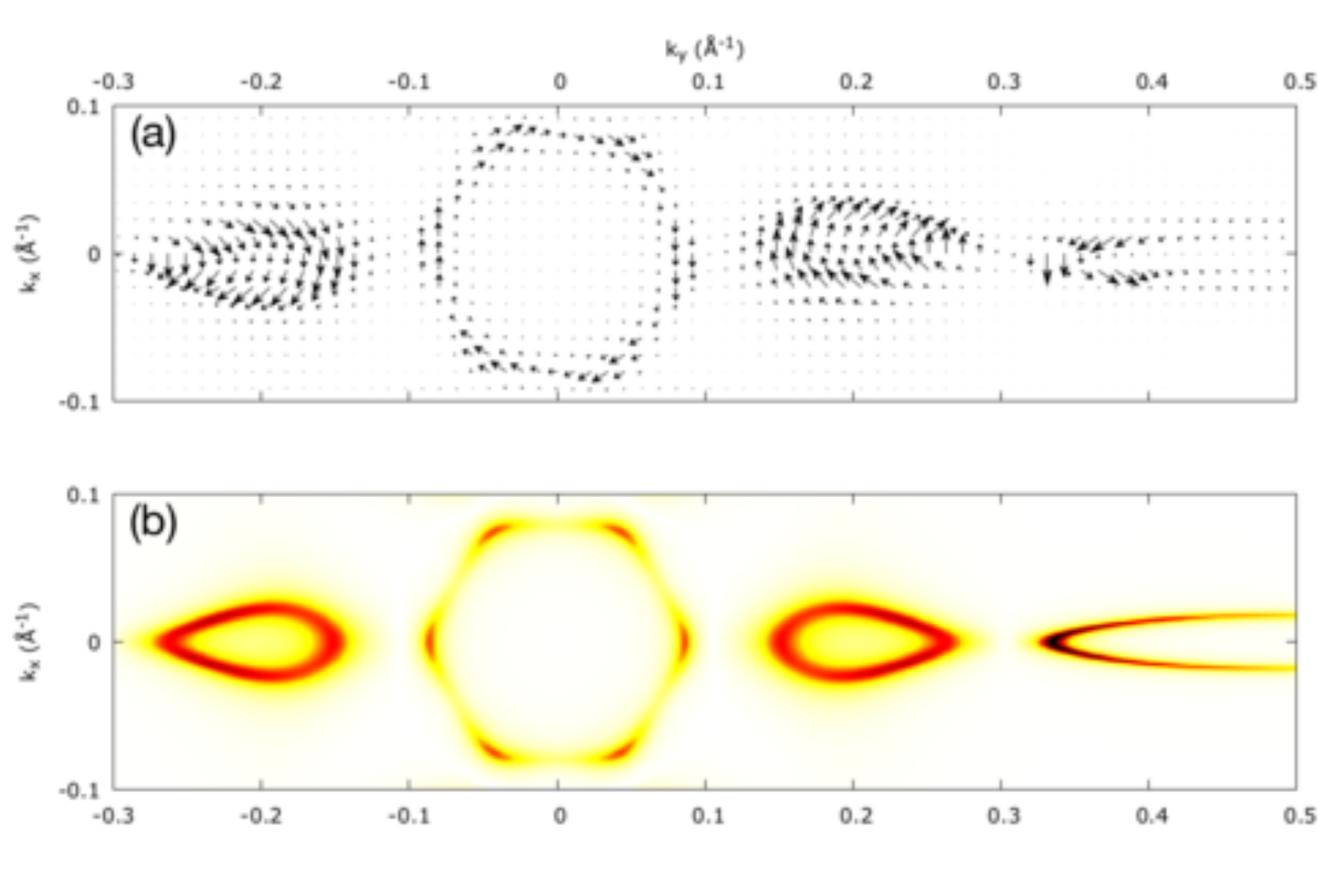}
\caption{%
(a) Magnitude and direction of $\vec{s}_{\shortparallel}$ as functions of
the wave vectors $k_x$ and $k_y$.
The length of the arrow at each point is proportional to $|\vec{s}_{\shortparallel}|$.
(b) $|\vec{s}_{\shortparallel}|$ for reference.
The same as Fig. 2(a) in the main text without the magnitude scale and green arrows.
}
\label{fig:S1}
\end{figure}

Figure~\ref{fig:S1}(a) shows the detailed magnitude and direction of
$\vec{s_{\shortparallel}}$ and Fig.~\ref{fig:S1}(b) shows a plot of $ |\vec{s_{\shortparallel}}|$
for reference.
The green arrows in Fig.~2(a) of the main text represent the directions of $\vec{s}_{\shortparallel}$ shown in Fig.~\ref{fig:S1}(a) as representative points.

\section{Surface spin DOS at the Fermi energy} \label{sec:}
\begin{figure}[htb]
\centering
\includegraphics[width=\textwidth]{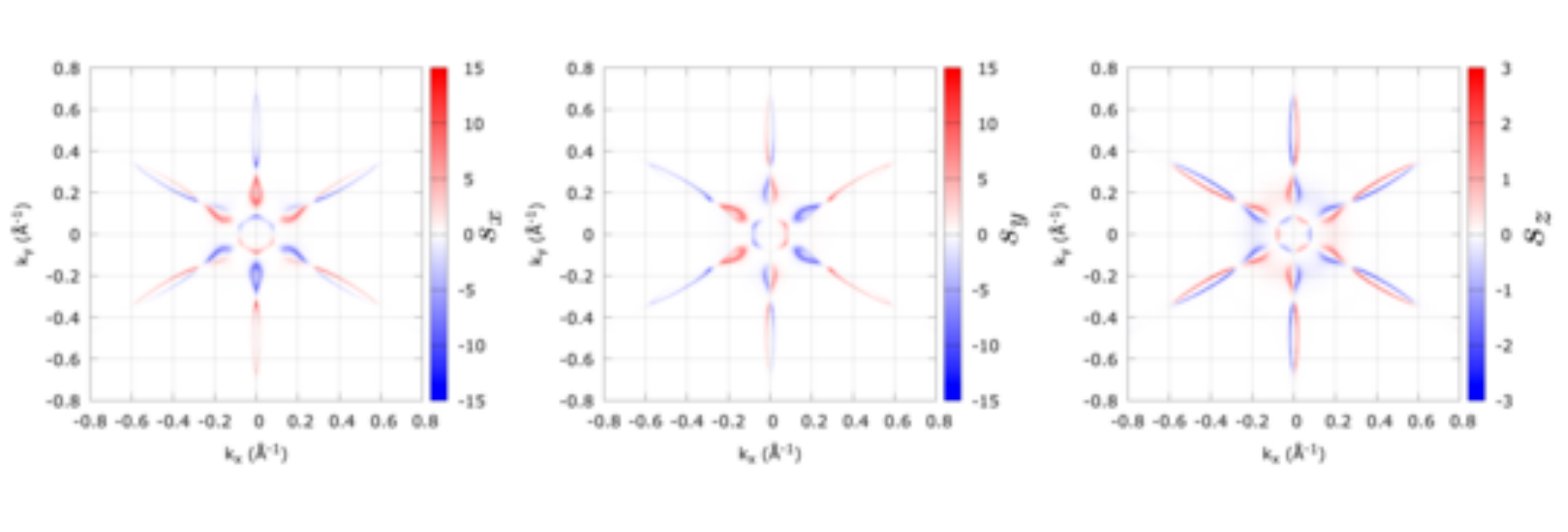}
\caption{%
Surface spin DOS 
at the Fermi energy:
$s_\alpha \equiv  s_\alpha(1,0,k_x,k_y)$
with $\alpha=x,y,z$.
}
\label{fig:S2}
\end{figure}

Figure~\ref{fig:S2} shows the surface spin DOS 
at the Fermi energy for all spin components.

\end{document}